\documentclass[11pt]{article}       
\usepackage{latexsym}               
\usepackage{url}                    
                                 {\null\hfill$\Box$\par\medskip}
\newtheorem{algX}{Algorithm}
\newenvironment{algorithm}       {\begin{algX}\begin{em}}%
                                 {\par\noindent --- End of Algorithm ---
                                 \end{em}\end{algX}}

\usepackage{graphicx}

\begin{document}


\title{{\bf Deterministic Sample Sort For GPUs}%
\thanks{Research partially supported by the Natural Sciences and Engineering
Research Council of Canada (NSERC).%
}}

\author{
{\bf Frank~Dehne}\\
{\small School of Computer Science}\\
{\small Carleton University}\\
{\small Ottawa, Canada K1S 5B6}\\
{\small {\em frank@dehne.net}}\\
{\small {\em http://www.dehne.net}}
\and 
{\bf Hamidreza Zaboli}\\
{\small School of Computer Science}\\
{\small Carleton University}\\
{\small Ottawa, Canada K1S 5B6}\\
{\small {\em hzaboli@connect.carleton.ca}}
} 

\maketitle

\begin{abstract}
We present and evaluate \textsc{GPU Bucket Sort}, a parallel \emph{deterministic}
sample sort algorithm for many-core GPUs. Our method is considerably
faster than Thrust Merge (Satish et.al., Proc. IPDPS 2009), the best
comparison-based sorting algorithm for GPUs, and it is as fast as
the new \emph{randomized} sample sort for GPUs by Leischner et.al.
(to appear in Proc. IPDPS 2010). Our \emph{deterministic} sample sort
has the advantage that bucket sizes are guaranteed and therefore its
running time does not have the input data dependent fluctuations that
can occur for randomized sample sort. 
\end{abstract}


\section{Introduction\label{sec:Introduction}}

Modern graphics processors (\emph{GPU}s) have evolved into highly
parallel and fully programmable architectures. Current many-core GPUs
such as nVIDIA's GTX and Tesla GPUs can contain up to 240 processor
cores on one chip and can have an astounding peak performance of up
to 1 TFLOP. The upcoming Fermi GPU recently announced by nVIDIA is
expected to have more than 500 processor cores. However, GPUs are
known to be hard to program and current general purpose (i.e. non-graphics)
GPU applications concentrate typically on problems that can be solved
using fixed and/or regular data access patterns such as image processing,
linear algebra, physics simulation, signal processing and scientific
computing (see e.g. \cite{gpgpu}). The design of efficient GPU methods
for discrete and combinatorial problems with data dependent memory
access patterns is still in its infancy. In fact, there is currently
still a lively debate even on the best \emph{sorting} method for GPUs
(e.g. \cite{s6-Garland,s7-Sanders}). 

Until very recently, the comparison-based Thrust Merge method \cite{s6-Garland}
by Nadathur Satish, Mark Harris and Michael Garland of nVIDIA Corporation
was considered the best sorting method for GPUs. However, an upcoming
paper by Nikolaj Leischner, Vitaly Osipov and Peter Sanders \cite{s7-Sanders}
(to appear in Proc. IPDPS 2010) presents a randomized sample sort
method for GPUs that significantly outperforms Thrust Merge. A disadvantage
of the randomized sample sort method is that its performance can vary
with the input data distribution because the data is partitioned into
buckets that are created via \emph{randomly} selected data items. 

In this paper, we present and evaluate a \emph{deterministic} sample
sort algorithm for GPUs, called \textsc{GPU Bucket Sort}, which has
the same performance as the randomized sample sort method in \cite{s7-Sanders}.
An experimental performance comparison on nVIDIA's GTX 285 and Tesla
architectures shows that for uniform data distribution, the\emph{
}best case\emph{ }for randomized sample sort, our deterministic sample
sort method is in fact \emph{exactly} as fast as the randomized sample
sort method of \cite{s7-Sanders}. However, in contrast to \cite{s7-Sanders},
the performance of \textsc{GPU Bucket Sort} remains the same for any
input data distribution because buckets are created deterministically
and bucket sizes are guaranteed.

The remainder of this paper is organized as follows. Section \ref{sec:Review:-GPU-Architectures}
reviews the \emph{Tesla} architecture framework for GPUs and the CUDA
programming environment, and Section \ref{sec:Previous-Work} reviews
previous work on GPU based sorting. Section \ref{sec:Deterministic-Sample-Sort}
presents \textsc{GPU Bucket Sort} and discusses some details of our
CUDA \cite{cuda-prog-guide} implementation. In Section \ref{sec:Experimental-Results-And},
we present an experimental performance comparison between our deterministic
sample sort implementation, the randomized sample sort implementation
in \cite{s7-Sanders}, and the Thrust Merge implementation in \cite{s6-Garland}.
In addition to the performance improvement discussed above, our deterministic
sample sort implementation appears to be more memory efficient as
well because \textsc{GPU Bucket Sort} is able to sort considerably
larger data sets within the same memory limits of the GPUs.

\section{Review: GPU Architectures\label{sec:Review:-GPU-Architectures}}

\begin{figure*}[tbh]
\begin{centering}
\includegraphics[width=12cm]{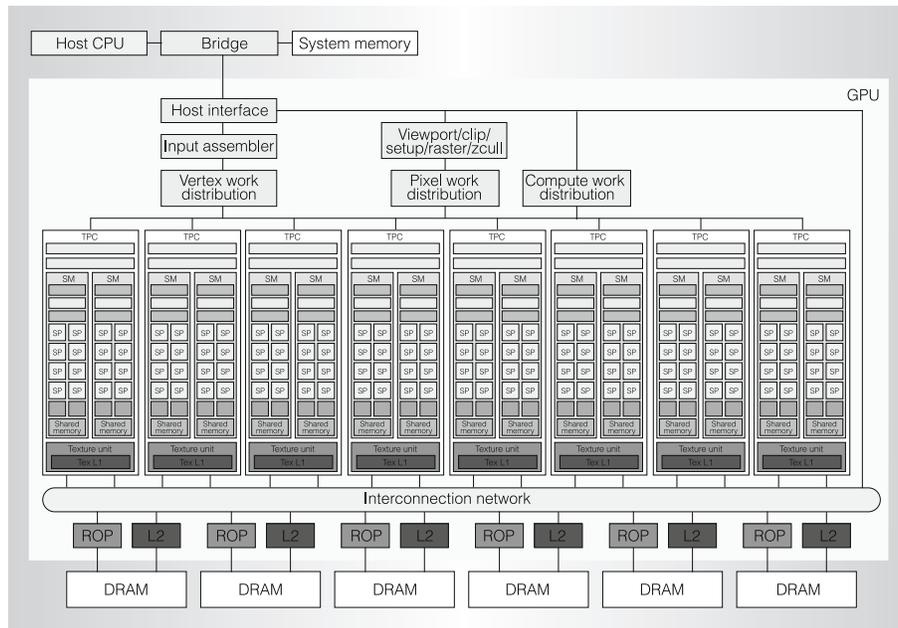}
\par\end{centering}

\centering{}\caption{\label{fig:nVIDIA-Tesla-Architecture}nVIDIA Tesla Architecture (from\cite{Lindholm2008})}

\end{figure*}

As in \cite{s6-Garland} and \cite{s7-Sanders}, we will focus on
nVIDIA's unified graphics and computing platform for GPUs known as
the \emph{Tesla} architecture framework \cite{Lindholm2008} and associated
\emph{CUDA} programming model \cite{cuda-prog-guide}. However, the
discussion and methods presented in this paper apply in general to
GPUs that support the OpenCL standard \cite{opencl-spec} which is
very similar to CUDA. A schematic diagram of the Tesla unified GPU
architecture is shown in Figure \ref{fig:nVIDIA-Tesla-Architecture}.
A Tesla GPU consists of an array of streaming processors called \emph{Streaming
Multiprocessors} \emph{(SM}s). Each SM contains eight processor cores
and a small size (16 KB) low latency local \emph{shared memory} that
is shared by its eight processor cores. All SMs are connected to a
\emph{global }DRAM\emph{ memory} through an interconnection network.
For example, an nVIDIA GeForce GTX 260 has 27 SMs with a total of
216 processor cores while GTX 285 and Tesla GPUs have 30 SMs with
a total of 240 processor cores. A GTX 260 has approximately 900 MB
global DRAM memory while GTX 285 and Tesla GPUs have up to 2 GB and
4 GB global DRAM memory, respectively (see Table \ref{tab:GPU-Performance-Characteristics}).
A GPU's global DRAM memory is arranged in independent memory partitions.
The interconnection network routes the read/write memory requests
from the processor cores to the respective global memory partitions,
and the results back to the cores. Each global memory partition has
its own queue for memory requests and arbitrates among the incoming
read/write requests, seeking to maximize DRAM transfer efficiency
by grouping read/write accesses to neighboring memory locations. Memory
latency to global DRAM memory is optimized when parallel read/write
operations can be grouped into a minimum number of arrays of contiguous
memory locations. 

It is important to note that data accesses from processor cores to
their SM's local shared memory are at least an order of magnitude
faster than accesses to global memory. This is an important consideration
for any efficient sorting method. Another critical issue for the performance
of CUDA implementations is conditional branching. CUDA programs typically
execute very large numbers of threads. In fact, a large number of
threads is critical for hiding latencies for global memory accesses.
The GPU has a hardware thread scheduler that is built to manage tens
of thousands and even millions of concurrent threads. All threads
are divided into blocks of up to 512 threads, and each block is executed
by an SM. An SM executes a thread block by breaking it into groups
of 32 threads called \emph{warps} and executing them in parallel using
its eight cores. These eight cores share various hardware components,
including the instruction decoder. Therefore, the threads of a warp
are executed in SIMT (single instruction, multiple threads) mode,
which is a slightly more flexible version of the standard SIMD (single
instruction, multiple data) mode. The main problem arises when the
threads encounter a conditional branch such as an IF-THEN-ELSE statement.
Depending on their data, some threads may want to execute the code
associated with the "true" condition and
some threads may want to execute the code associated with the "false"
condition. Since the shared instruction decoder can only handle one
branch at a time, different threads can not execute different branches
concurrently. They have to be executed in sequence, leading to performance
degradation. GPUs provide a small improvement through an instruction
cache at each SM that is shared by its eight cores. This allows for
a "small" deviation between the instructions
carried out by the different cores. For example, if an IF-THEN-ELSE
statement is short enough so that both conditional branches fit into
the instruction cache then both branches can be executed fully in
parallel. However, a poorly designed algorithm with too many and/or
large conditional branches can result in serial execution and very
low performance.

\section{Previous Work On GPU Sorting\label{sec:Previous-Work}}

Sorting algorithms for GPUs started to appear a few years ago and
have been highly competitive. Early results include GPUTeraSort \cite{s1-GPUTeraSort}
based on bitonic merge, and adaptive bitonic sort \cite{s2-GPU-ABiSort}
based on a method by Bilardi et.al. \cite{s3-Bilardi}. Hybrid sort
\cite{s4-hybrid} used a combination of bucket sort and merge sort,
and D. Cederman et.al. \cite{s5-practical-quicksort} proposed a quick
sort based method for GPUs. Both methods (\cite{s4-hybrid,s5-practical-quicksort})
suffer from load balancing problems. Until very recently, the comparison-based
Thrust Merge method \cite{s6-Garland} by Nadathur Satish, Mark Harris
and Michael Garland of nVIDIA Corporation was considered the best
sorting method for GPUs. Thrust Merge uses a combination of odd-even
merge and two-way merge, and overcomes the load balancing problems
mentioned above. Satish et.al. \cite{s6-Garland} also presented an
even faster GPU radix sort method for the special case of integer
sorting. Yet, an upcoming paper by Nikolaj Leischner, Vitaly Osipov
and Peter Sanders \cite{s7-Sanders} (to appear in Proc. IPDPS 2010)
presents a randomized sample sort method for GPUs that significantly
outperforms Thrust Merge \cite{s6-Garland}. However, as discussed
in Section \ref{sec:Introduction}, the performance of randomized
sample sort can vary with the distribution of the input data because
buckets are created through randomly selected data items. Indeed,
the performance analysis presented in \cite{s7-Sanders} measures
the runtime of their randomized sample sort method for six different
data distributions to document the performance variations observed
for different input distributions.

\section{\textsc{GPU Bucket Sort}\emph{:\\Deterministic} Sample Sort For
GPUs\label{sec:Deterministic-Sample-Sort}}

In this section we present \textsc{GPU Bucket Sort}, a \emph{deterministic}
sample sort algorithm for GPUs, and discuss its CUDA implementation.
An outline of \textsc{GPU Bucket Sort} is shown in Algorithm \ref{alg:Deterministic-Sample-Sort}
below. It consists of a local sort (Step 1), a selection of samples
that define balanced buckets (Steps 3-5), moving all data into those
buckets (Steps 6-8), and a final sort of each bucket. 


\begin{algorithm}\label{alg:Deterministic-Sample-Sort}
\textsc{GPU Bucket Sort} (Deterministic Sample Sort For GPUs)

\emph{Input}: An array $A$ with $n$ data items stored in global
memory.

\emph{Output}: Array $A$ sorted.
\begin{enumerate}
\item Split the array $A$ into $m$ sublists $A_{1},...,A_{m}$ containing
$\frac{n}{m}$ items each where $\frac{n}{m}$ is the shared memory
size at each SM.
\item \emph{Local Sort}: Sort each sublist $A_{i}$ ($i$=1,..., $m$) locally
on one SM, using the SM's shared memory as a cache.
\item \emph{Local Sampling}: Select $s$ equidistant samples from each sorted
sublist $A_{i}$ ($i$=1,..., $m$) for a total of $sm$ samples.
\item \emph{Sorting All Samples}: Sort all $sm$ samples in global memory,
using all available SMs in parallel. 
\item \emph{Global Sampling}: Select $s$ equidistant samples from the sorted
list of $sm$ samples. We will refer to these $s$ samples as \emph{global
samples}.
\item \emph{Sample Indexing}: For each sorted sublist $A_{i}$ ($i$=1,...,
$m$) determine the location of each of the $s$ global samples in
$A_{i}$. This operation is done for each $A_{i}$ locally on one
SM, using the SM's shared memory, and will create for each $A_{i}$
a partitioning into $s$ buckets $A_{i1}$,..., $A_{is}$ of size
$a_{i1}$,..., $a_{is}$.
\item \emph{Prefix Sum}: Through a parallel prefix sum operation on $a_{11}$,...,
$a_{m1}$, $a_{12}$,..., $a_{m2}$, ..., $a_{1s}$,..., $a_{ms}$
calculate for each bucket $A_{ij}$($1 \leq i \leq m$, $1 \leq j \leq s$,
) its starting location $l_{ij}$ in the final sorted sequence.
\item \emph{Data Relocation}: Move all $sm$ buckets $A_{ij}$($1\leq i \leq m$,
1$\leq j \leq s$) to location $l_{ij}$. The newly created
array consists of $s$ sublists $B_{1}$, ..., $B_{s}$ where 
$B_{j}=A_{1j}\cup A_{2j}\cup...\cup A_{mj}$
for 1$\leq j \leq s$.
\item \emph{Sublist Sort}: Sort all sublists $B_{j}$, 1$\leq j \leq s$,
using all SMs.
\end{enumerate}

\end{algorithm}

\begin{figure}[tbh]
\begin{centering}
\includegraphics[width=7cm]{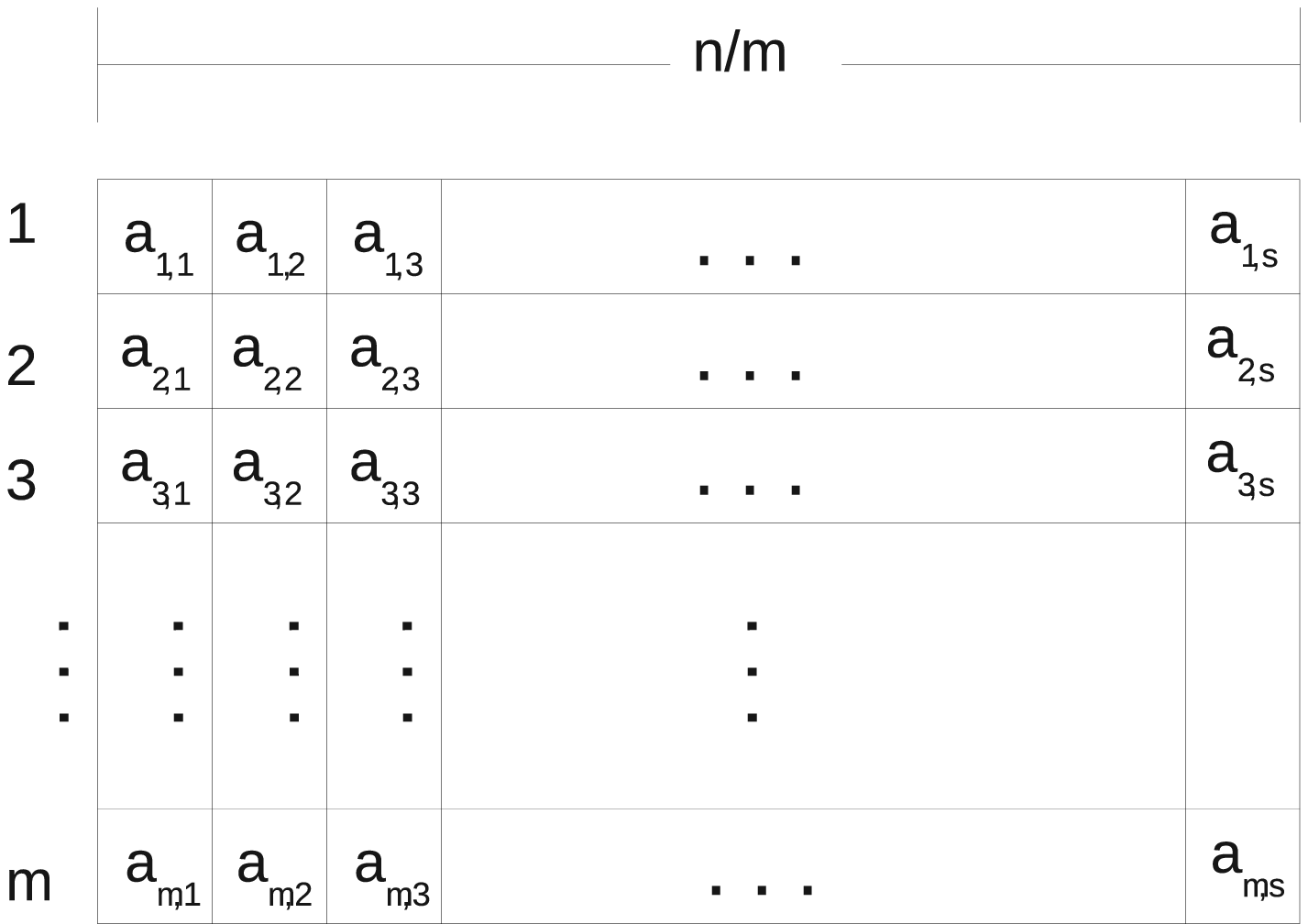}
\par\end{centering}

\caption{Illustration Of Step 7 In Algorithm \ref{alg:Deterministic-Sample-Sort}\label{fig:Illustration-Of-Step-7}}

\end{figure}

Our discussion of Algorithm \ref{alg:Deterministic-Sample-Sort} and
its implementation will focus on GPU performance issues related to
shared memory usage, coalesced global memory accesses, and avoidance
of conditional branching. Consider an input array $A$ with $n=32M$
data items and a local shared memory size of $\frac{n}{m}=2K$ data
items. In Steps 1 and 2 of Algorithm \ref{alg:Deterministic-Sample-Sort},
we split the array A into $m=16K$ sublists of $2K$ data items each
and then locally sort each of those $m=16K$ sublists. More precisely,
we create 16 K thread blocks of 512 threads each, where each thread
block sorts one sublist using one SM. Each thread block first loads
a sublist into the SM's local shared memory using a coalesced parallel
read from global memory. Note that, each of the 512 threads is responsible
for $\frac{n}{m}/512=4$ data items. The thread block then sorts a
sublist of $\frac{n}{m}=2K$ data items in the SM's local shared memory.
We tested different implementations for the local shared memory sort
within an SM, including quicksort, bitonic sort, and adaptive bitonic
sort \cite{s3-Bilardi}. In our experiments, bitonic sort was consistently
the fastest method, despite the fact that it requires $O(n\log^{2}n)$
work. The reason is that, for Step 2 of Algorithm \ref{alg:Deterministic-Sample-Sort},
we always sort $2K$ data items only, irrespective of $n$. For such
a small number of items the simplicity of bitonic sort, it's small
constants in the running time, and it's perfect match for SIMD style
parallelism outweigh the disadvantage of additional work. In Step
3 of Algorithm \ref{alg:Deterministic-Sample-Sort}, we select $s=64$
equidistant samples from each sorted sublist. The choice of value
for $s$ is discussed in Section \ref{sec:Experimental-Results-And}.
The implementation of Step 3 is built directly into the final phase
of Step 2 when the sorted sublists are written back into global memory.
In Step 4, we are sorting all $sm=1M$ selected samples in global
memory, using all available SMs in parallel. Here, we compared GPU
bitonic sort \cite{s1-GPUTeraSort}, adaptive bitonic sort \cite{s2-GPU-ABiSort}
based on \cite{s3-Bilardi}, and randomized sample sort \cite{s7-Sanders}.
Our experiments indicate that for up to 16 M data items, simple bitonic
sort is still faster than even randomized sample sort \cite{s7-Sanders}
due to its simplicity, small constants, and complete avoidance of
conditional branching. Hence, Step 4 was implemented via bitonic sort.
In Step 5, we select $s=64$ equidistant \emph{global samples} from
the sorted list of $sm=1M$ samples. Here, each thread block/SM loads
the $s=64$ global samples into its local shared memory where they
will remain for the next step. In Step 6, we determine for each sorted
sublist $A_{i}$ ($i$=1, ..., $m$) of $\frac{n}{m}=2K$ data items
the location of each of the $s=64$ global samples in $A_{i}$. For
each $A_{i}$, this operation is done locally by one thread block
on one SM, using the SM's shared memory, and will create for each
$A_{i}$ a partitioning into $s=64$ buckets $A_{i1}$,..., $A_{is}$
of size $a_{i1}$,..., $a_{is}$. Here, we apply parallel binary search
in $A_{i}$ for each of the global samples. More precisely, we first
take the $\frac{s}{2}$-th global sample element and use one thread
to perform a binary search in $A_{i}$, resulting in a location $l_{s/2}$
in $A_{i}$. Then we use two threads to perform two binary searches
in parallel, one for the $\frac{s}{4}$-th global sample element in
the part of $A_{i}$ to the left of location $l_{s/2}$ , and one
for the $\frac{3s}{4}$-th global sample element in the part of $A_{i}$
to the right of location $l_{s/2}$. This process is iterated $\log s=6$
times until all $s=64$ global samples are located in $A_{i}$. With
this, each $A_{i}$ is split into $s=64$ buckets $A_{i1}$,..., $A_{is}$
of size $a_{i1}$,..., $a_{is}$. Note that, we do not simply perform
all $s$ binary searches fully in parallel in order to avoid memory
contention within the local shared memory\cite{cuda-prog-guide}.
Step 7 uses a prefix sum calculation to obtain for all buckets their
starting location in the final sorted sequence. The operation is illustrated
in Figure \ref{fig:Illustration-Of-Step-7} and can be implemented
with coalesced memory accesses in global memory. Each row in Figure
\ref{fig:Illustration-Of-Step-7} shows the $a_{i1}$,..., $a_{is}$
calculated for each sublist. The prefix sum is implemented via a parallel
column sum (using all SMs), followed by a prefix sum on the columns
sums (on one SM in local shared memory), and a final update of the
partial sums in each column (using all SMs). In Step 8, the $sm=1M$
buckets are moved to their correct location in the final sorted sequence.
This operation is perfectly suited for a GPU and requires one parallel
coalesced data read followed by one parallel coalesced data write
operation. The newly created array consists of $s=64$ sublists $B_{1}$,
..., $B_{s}$ where each $B_{j}=A_{1j}\cup A_{2j}\cup...\cup A_{mj}$
has at most $\frac{2n}{s}=1M$ data items \cite{Schaeffer}. In Step
9, we sort each $B_{j}$ using the same bitonic sort implementation
as in Step 4. Note that, since each $B_{j}$ is smaller than $16M$
data items, simple bitonic sort is faster for each $B_{j}$ than even
randomized sample sort \cite{s7-Sanders} due to bitonic sort's simplicity,
small constants, and complete avoidance of conditional branching.

\section{Experimental Results And Discussion\label{sec:Experimental-Results-And}}

\begin{figure*}[tbh]
\begin{centering}
\includegraphics[width=13cm]{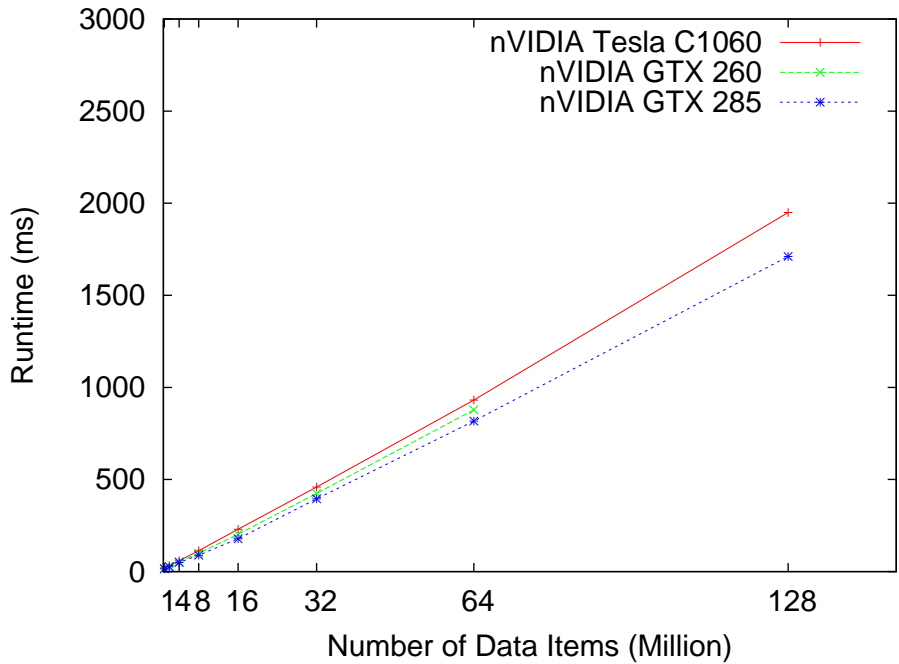} 
\par\end{centering}

\centering{}\caption{Performance Of Deterministic Sample Sort For GPUs (\textsc{GPU Bucket
Sort}). Total runtime for varying number of data items on different
GPUs: nVIDIA Tesla C1060, GTX 260 and GTX 285.\label{fig:det-260-285-tesla-comp}}

\end{figure*}

For our experimental evaluation, we executed Algorithm \ref{alg:Deterministic-Sample-Sort}
on three different GPUs (nVIDIA Tesla, GTX 285, and GTX 260) for various
data sets of different sizes, and compared our results with those
reported in \cite{s6-Garland} and \cite{s7-Sanders} which are the
current best GPU sorting methods. Table \ref{tab:GPU-Performance-Characteristics}
shows some important performance characteristics of the three different
GPUs. The Tesla and GTX 285 have more cores than the GTX 260. The
GTX 285 has the highest core clock rate and in summary the highest
level of core computational power. The Tesla has the largest memory
but the GTX 285 has the best memory clock rate and memory bandwidth.
In fact, even the GTX 260 has a higher clock rate and memory bandwidth
than the Tesla C1060. Figure \ref{fig:det-260-285-tesla-comp} shows
a comparison of the runtime of our \textsc{GPU Bucket Sort} method
on the Tesla C1060, GTX 260 and GTX 285 (with 2 GB memory) for varying
number of data items. Each data point shows the average of 100 experiments.
The observed variance was less than 1 ms for all data points since
\textsc{GPU Bucket Sort} is deterministic and any fluctuation observed
was due to noise on the GPU (e.g. operating system related traffic).
All three curves show a growth rate very close to linear which is
encouraging for a problem that requires $O(n\log n)$ work. \textsc{GPU
Bucket Sort} performs better on the GTX 285 than both Tesla and GTX
260. Furthermore, it performs better on the GTX 260 than on the Tesla
C1060. This indicates that \textsc{GPU Bucket Sort} is memory bandwidth
bound which is expected for sorting methods since the sorting problem
requires only very little computation but a large amount of data movement.
For individual steps of \textsc{GPU Bucket Sort}, the order can sometimes
be reversed. For example, we observed that Step 2 of Algorithm \ref{alg:Deterministic-Sample-Sort}
(local sort) runs faster on the Tesla C1060 than on the GTX 260 since
this step is executed locally on each SM and its performance is largely
determined by the number of SMs and the performance of the SM's cores.
However, the GTX 285 remained the fastest machine, even for all individual
steps.

\begin{figure*}[tbh]
\begin{centering}
\includegraphics[width=13cm]{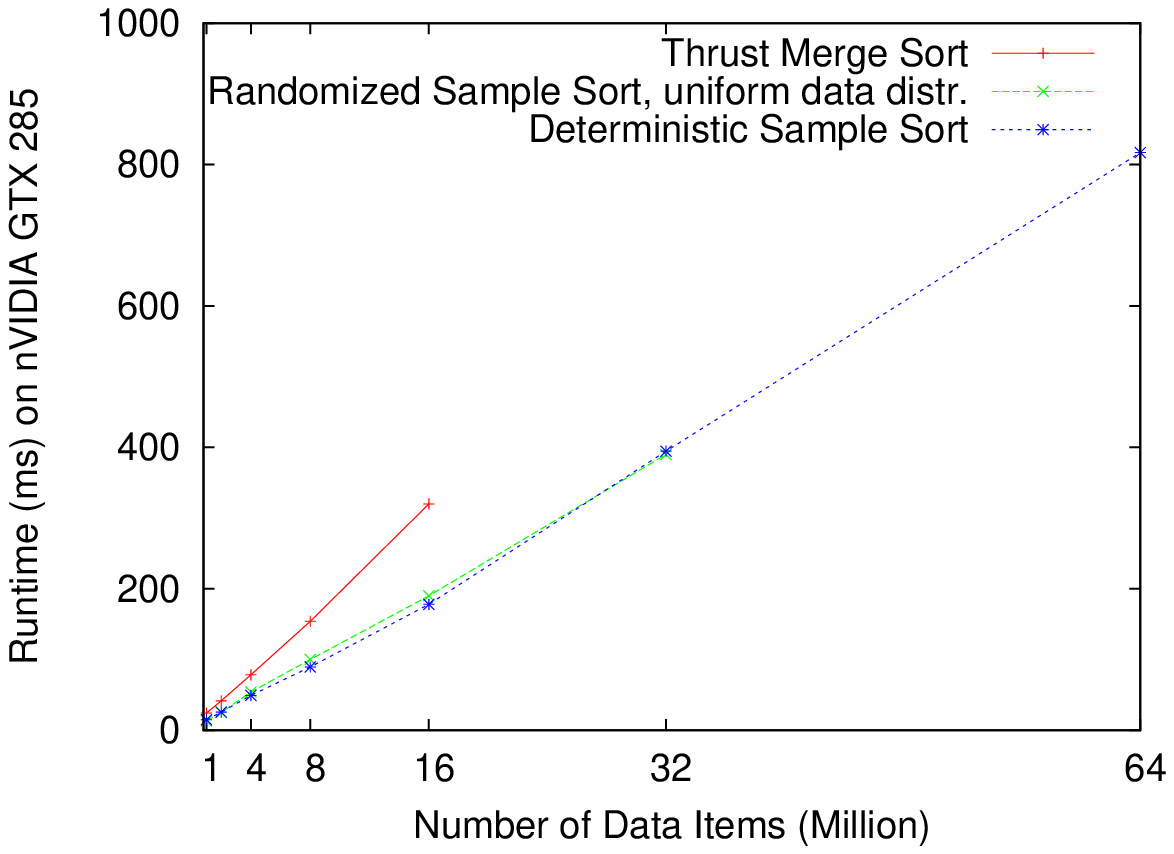}
\par\end{centering}

\begin{centering}
(a) Number of Data Items Up To 64,000,000.
\par\end{centering}

\begin{centering}
\includegraphics[width=13cm]{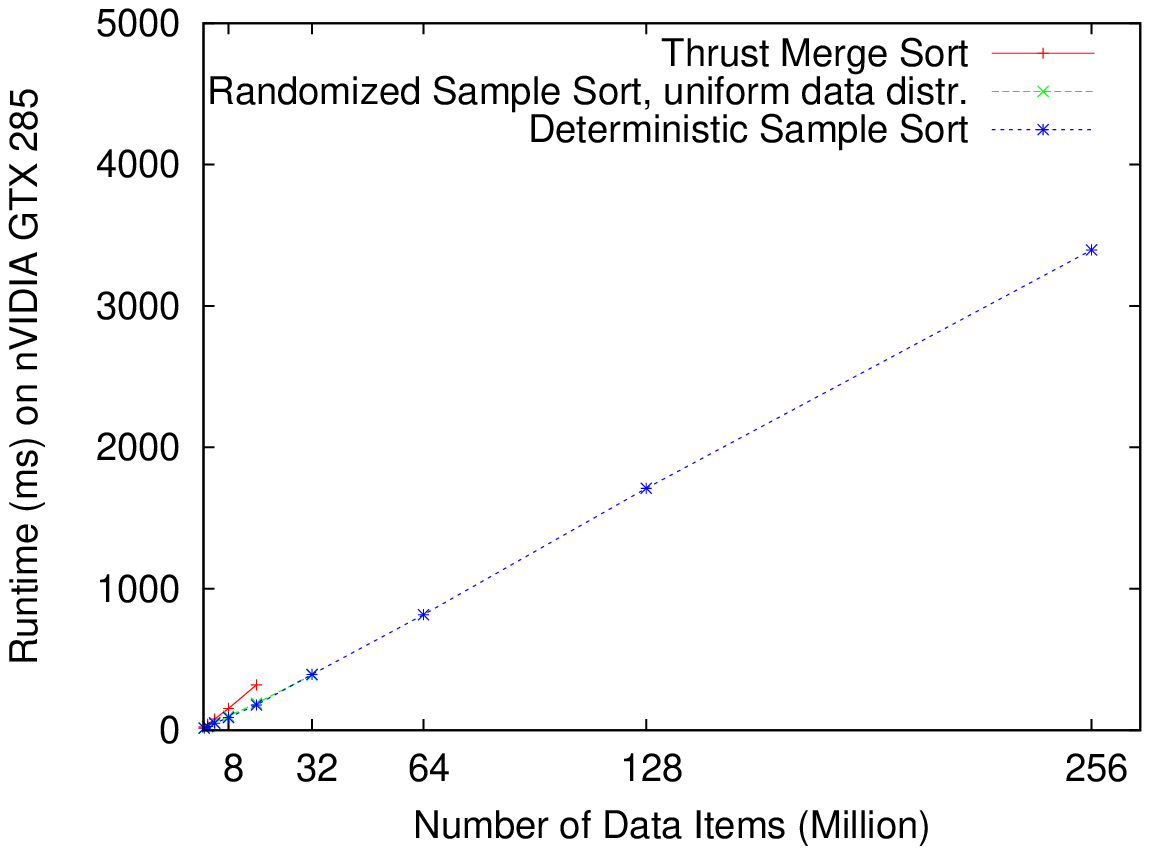}
\par\end{centering}

\begin{centering}
(b) Number of Data Items Up To 512,000,000.
\par\end{centering}

\centering{}\caption{Comparison between Deterministic Sample Sort (\textsc{GPU Bucket Sort}),
Randomized Sample Sort \cite{s7-Sanders} and Thrust Merge Sort \cite{s6-Garland}.
Total runtime for varying number of data items on an nVIDIA GTX 285.
(\cite{s6-Garland} and \cite{s7-Sanders} provided data only for
up to 16M and 32M data items, respectively.)\label{fig:det-rand-merge-285}}

\end{figure*}

\begin{figure*}[tbh]
\begin{centering}
\includegraphics[width=13cm]{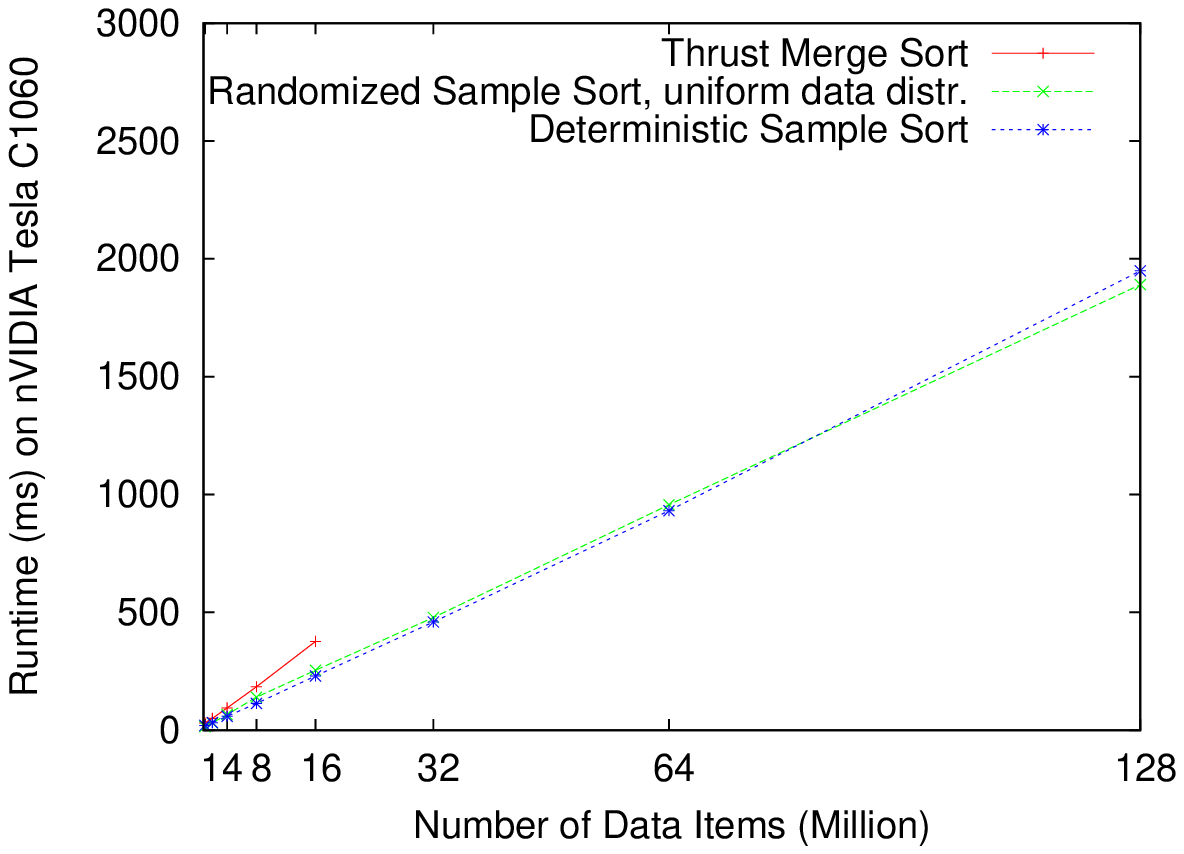}
\par\end{centering}

\begin{centering}
(a) Number of Data Items Up To 128,000,000.
\par\end{centering}

\begin{centering}
\includegraphics[width=13cm]{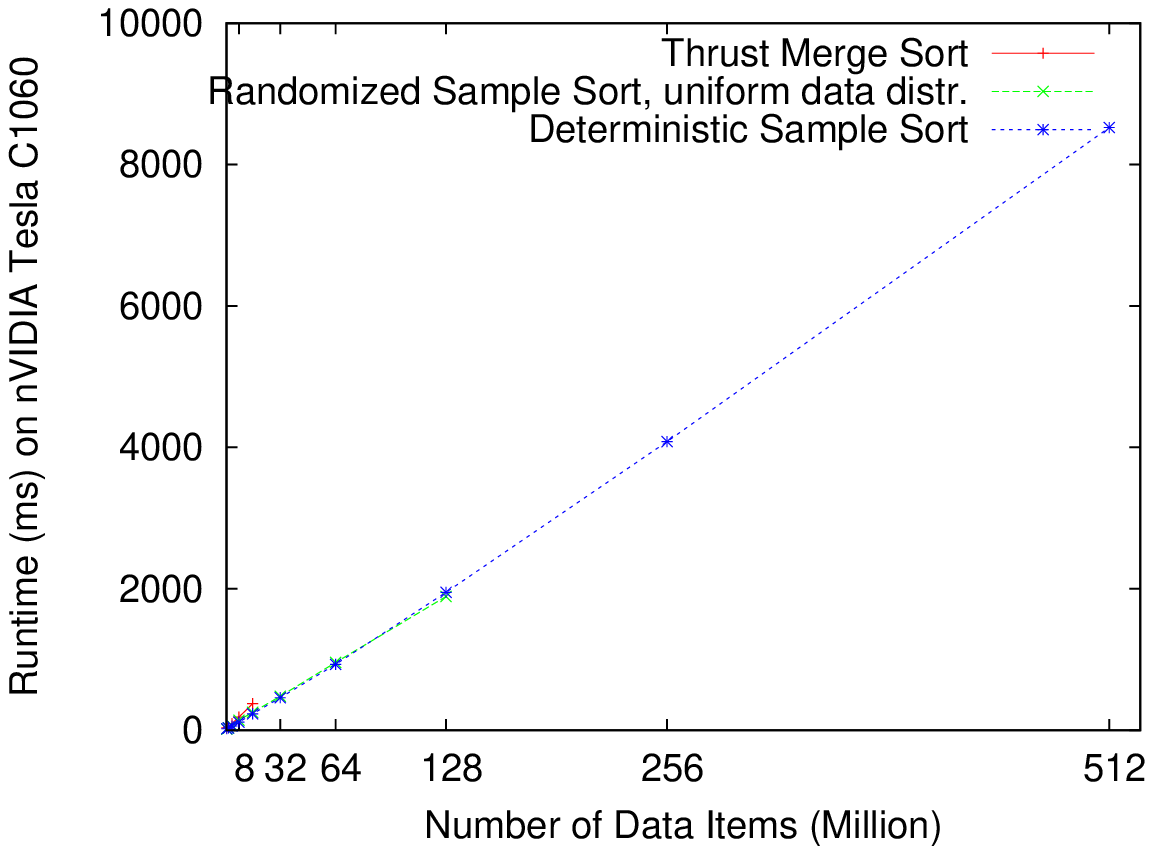}
\par\end{centering}

\begin{centering}
(b) Number of Data Items Up To 512,000,000.
\par\end{centering}

\centering{}\caption{Comparison between Deterministic Sample Sort (\textsc{GPU Bucket Sort}),
Randomized Sample Sort \cite{s7-Sanders} and Thrust Merge Sort \cite{s6-Garland}.
Total runtime for varying number of data items on an nVIDIA Tesla
C1060.(\cite{s6-Garland} and \cite{s7-Sanders} provided data only
for up to 16M and 128M data items, respectively.) \label{fig:det-rand-merge-tesla} }

\end{figure*}

We note that \textsc{GPU Bucket Sort} can sort up to $n=64M$ data
items within the 896 MB memory available on the GTX 260 (see Figure
\ref{fig:det-260-285-tesla-comp}). On the GTX 285 with 2 GB memory
and Tesla C1060 our \textsc{GPU Bucket Sort} method can sort up to
$n=256M$ and $n=512M$ data items, respectively (see Figures \ref{fig:det-rand-merge-285}\&\ref{fig:det-rand-merge-tesla}).

\begin{figure*}[tbh]
\centering{}\includegraphics[width=5cm]{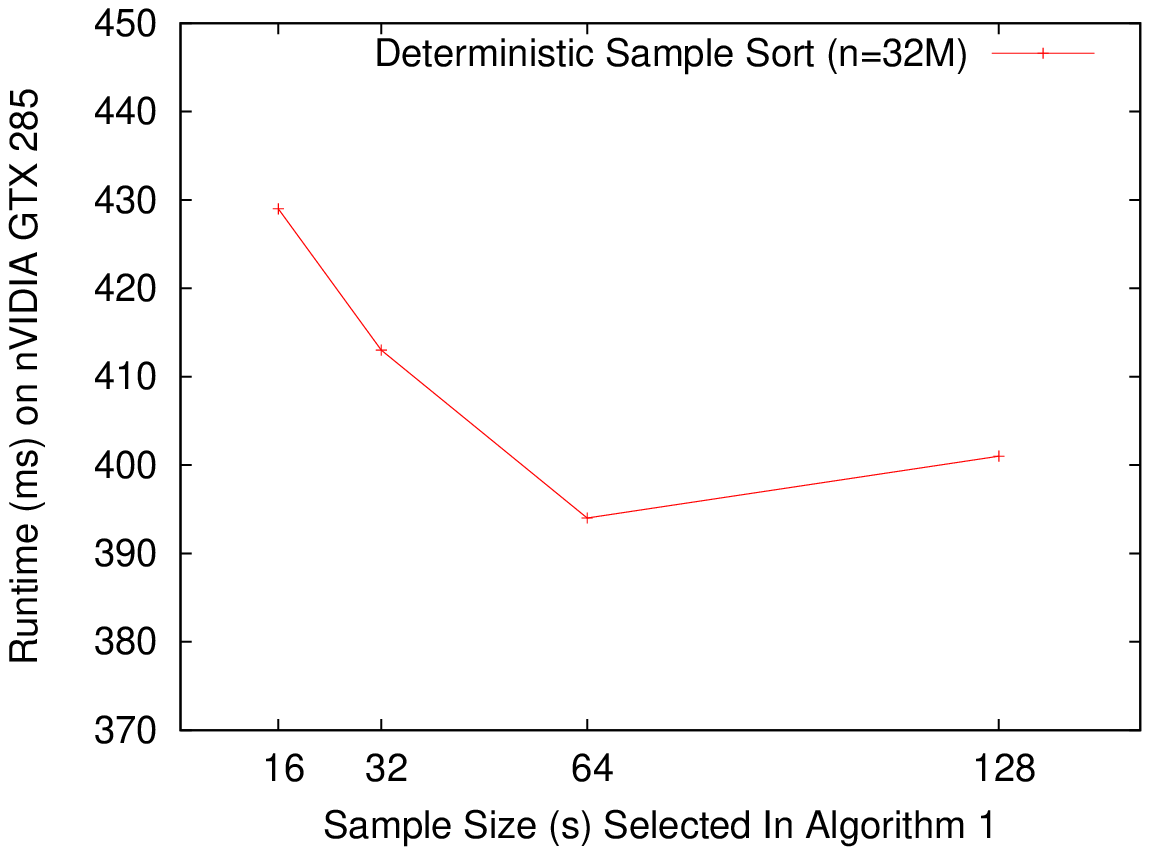}
\includegraphics[width=5cm]{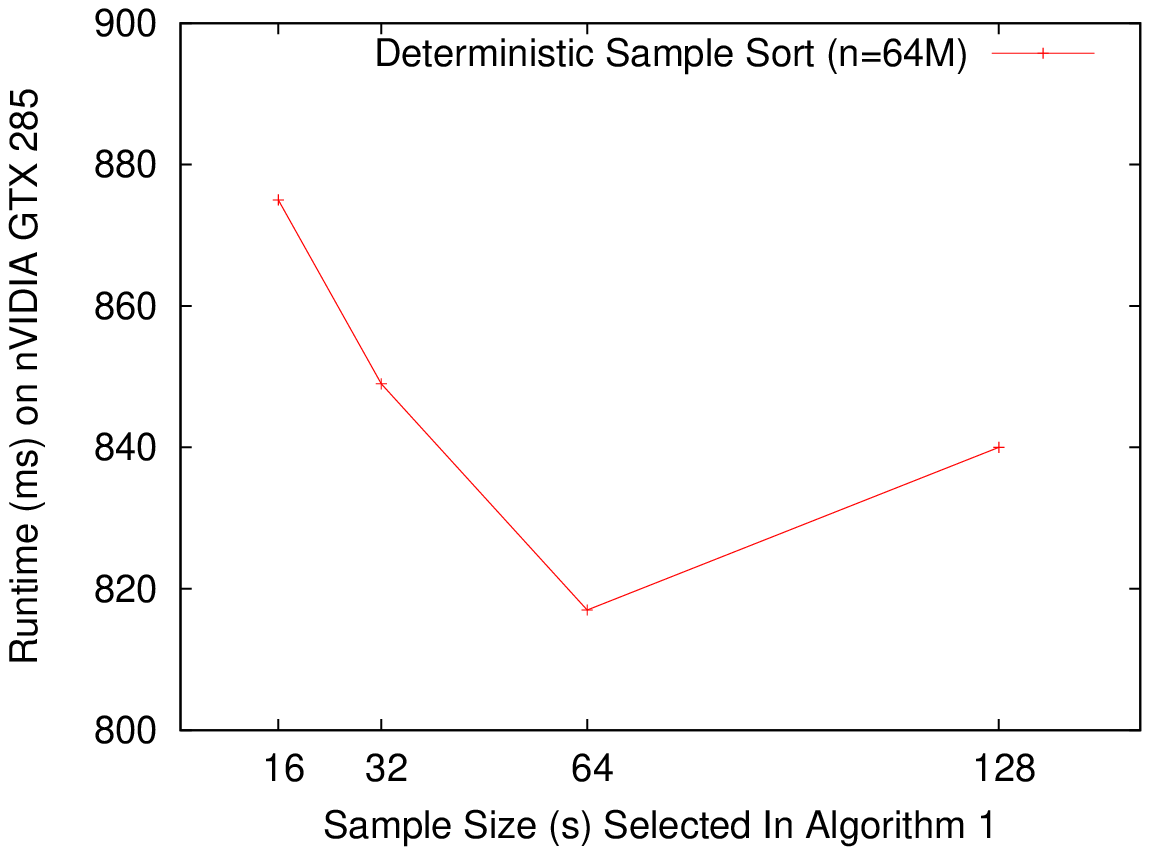}
\includegraphics[width=5cm]{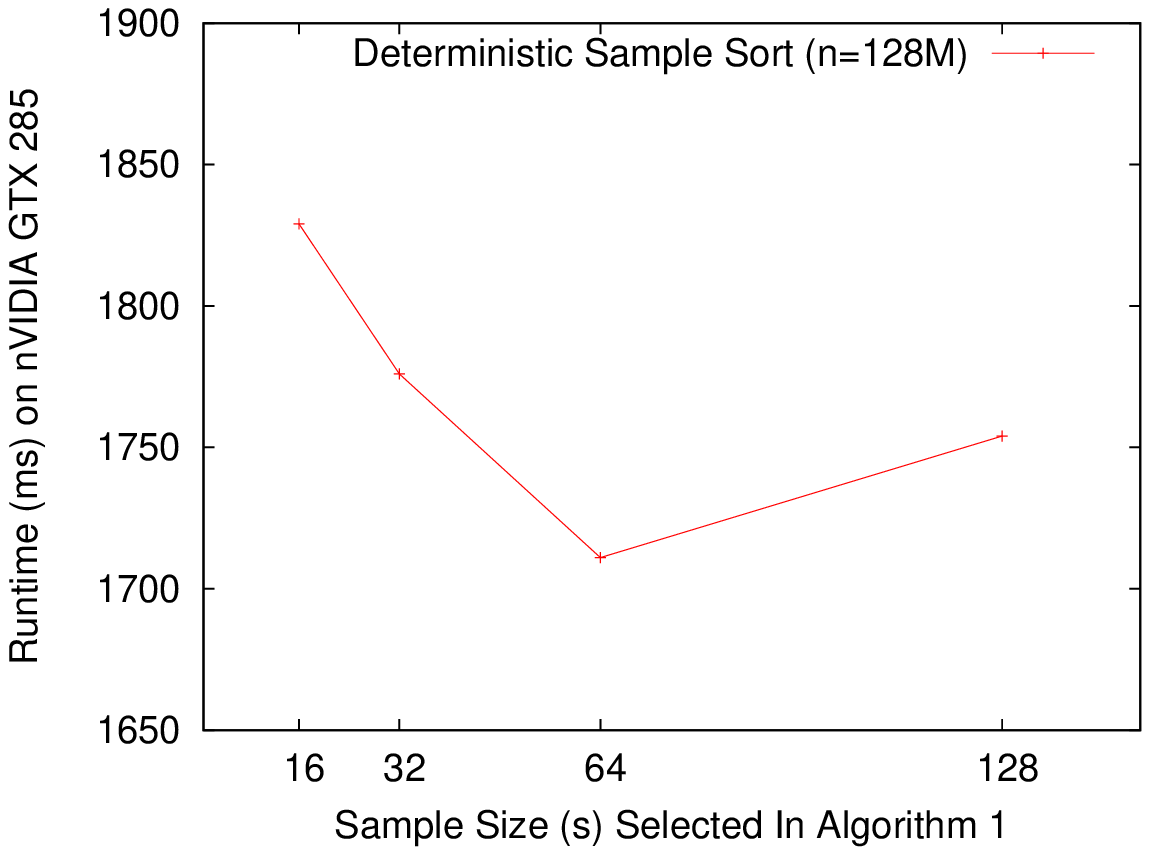}\caption{Runtime Of Algorithm \ref{alg:Deterministic-Sample-Sort} As A Function
Of Selected Sample Size $s$ For Fixed $n=32M$, $n=64M$, and $n=128M$.
\label{fig:Runtime-Sample-Size}}

\end{figure*}

\begin{figure*}[tbh]
\begin{centering}
\includegraphics[width=13cm]{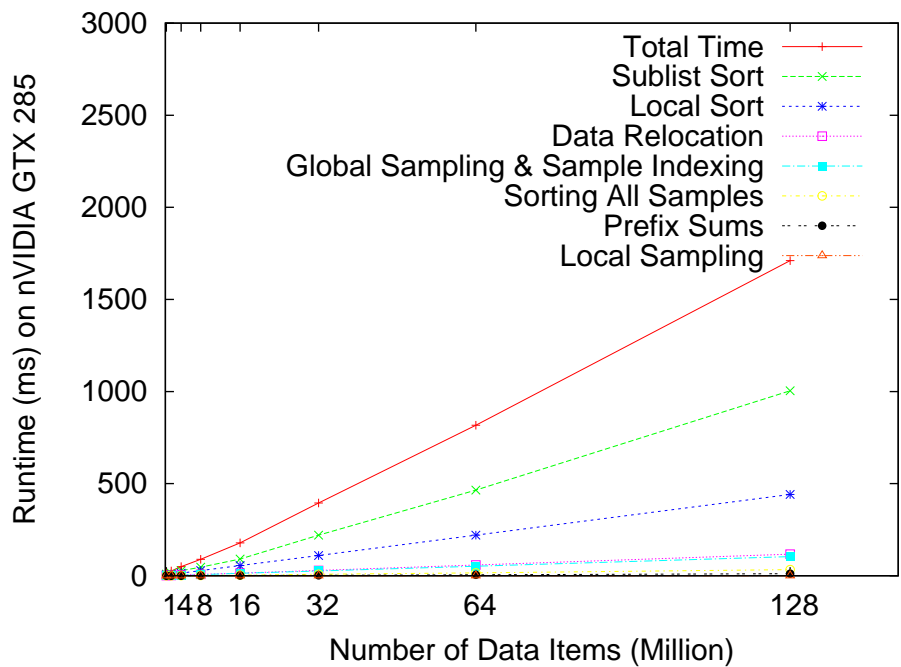}
\par\end{centering}

\centering{}\caption{Performance Of Deterministic Sample Sort For GPUs (\textsc{GPU Bucket
Sort}). Total runtime and runtime for individual steps of Algorithm
\ref{alg:Deterministic-Sample-Sort} on an nVIDIA GTX 285 for varying
number of data items.\label{fig:det-steps}}

\end{figure*}

Figure \ref{fig:det-steps} shows in detail the time required for
the individual steps of Algorithm \ref{alg:Deterministic-Sample-Sort}
when executed on a GTX 285. We observe that \emph{sublist sort} (Step
9) and \emph{local sort} (Step 2) represent the largest portion of
the total runtime of \textsc{GPU Bucket Sort}. This is very encouraging
in that the {}``overhead'' involved to manage the deterministic
sampling and generate buckets of guaranteed size (Steps 3-7) is small.
We also observe that the \emph{data relocation} operation (Step 8)
is very efficient and a good example of the GPU's great performance
for data parallel access when memory accesses can be coalesced (see
Section \ref{sec:Review:-GPU-Architectures}). Note that, for Algorithm
\ref{alg:Deterministic-Sample-Sort} the sample size $s$ is a free
parameter. With increasing $s$, the sizes of sublists $B_{j}$ created
in Step 8 of Algorithm \ref{alg:Deterministic-Sample-Sort} decrease
and the time for Step 9 decreases as well. However, the time for Steps
3-7 grows with increasing $s$. This trade-off is illustrated in Figure
\ref{fig:Runtime-Sample-Size} which shows the total runtime for Algorithm
\ref{alg:Deterministic-Sample-Sort} as a function of $s$ for fixed
$n=32M,64M,128M$. As shown in Figure \ref{fig:Runtime-Sample-Size},
the total runtime is smallest for $s=64$, which is the parameter
chosen for our \textsc{GPU Bucket Sort} code.

Figures \ref{fig:det-rand-merge-285} and \ref{fig:det-rand-merge-tesla}
show a comparison between \textsc{GPU Bucket Sort} and the current
best GPU sorting methods, Randomized Sample Sort \cite{s7-Sanders}
and Thrust Merge Sort \cite{s6-Garland}. Figure \ref{fig:det-rand-merge-285}
shows the runtimes for all three methods on a GTX 285 and Figure \ref{fig:det-rand-merge-tesla}
shows the runtimes of all three methods on a Tesla C1060. Note that,
\cite{s6-Garland} and \cite{s7-Sanders} did not report runtimes
for the GTX 260. For \textsc{GPU Bucket Sort}, all runtimes are the
averages of 100 experiments, with less than 1 ms observed variance.
For Randomized Sample Sort and Thrust Merge Sort, the runtimes shown
are the ones reported in \cite{s7-Sanders} and \cite{s6-Garland}.
For Thrust Merge Sort, performance data is only available for up to
$n=16M$ data items. For larger values of $n$, the current Thrust
Merge Sort code shows memory errors \cite{Garland-Priv-Comm}. As
reported in \cite{s7-Sanders}, the current Randomized Sample Sort
code can sort up to $32M$ data items on a GTX 285 with 1 GB memory
and up to $128M$ data items on a Tesla C1060. Our \textsc{GPU Bucket
Sort} code appears to be more memory efficient. \textsc{GPU Bucket
Sort }can sort up to $n=256M$ data items on a GTX 285 with 2GB memory
and up to $n=512M$ data items on a Tesla C1060. Therefore, Figures
\ref{fig:det-rand-merge-285}a and \ref{fig:det-rand-merge-tesla}a
show the performance comparison with higher resolution for up to $n=64M$
and $n=128M$, respectively, while Figures \ref{fig:det-rand-merge-285}b
and \ref{fig:det-rand-merge-tesla}b show the performance comparison
for the entire range up to $n=256M$ and $n=512M$, respectively. 

We observe in Figures \ref{fig:det-rand-merge-285}a and \ref{fig:det-rand-merge-tesla}a
that, as reported in \cite{s7-Sanders}, Randomized Sample Sort \cite{s7-Sanders}
significantly outperforms Thrust Merge Sort \cite{s6-Garland}. Most
importantly, we observe that Randomized Sample Sort \cite{s7-Sanders}
and our Deterministic Sample Sort (\textsc{GPU Bucket Sort}) show
nearly identical performance on both, the GTX 285 and Tesla C1060.
Note that, the experiments in \cite{s7-Sanders} used a GTX 285 with
1 GB memory whereas we used a GTX 285 with 2 GB memory. As shown in
Table \ref{tab:GPU-Performance-Characteristics}, the GTX 285 with
1 GB has a slightly better memory clock rate and memory bandwidth
than the GTX 285 with 2 GB which implies that the performance of Deterministic
Sample Sort (\textsc{GPU Bucket Sort}) on a GTX 285 is actually a
few percent better than the performance of Randomized Sample Sort.
The data sets used for the performance comparison in Figures \ref{fig:det-rand-merge-285}
and \ref{fig:det-rand-merge-tesla} were uniformly distributed, random
data items. The data distribution does not impact the performance
of Deterministic Sample Sort (\textsc{GPU Bucket Sort}) but has an
impact on the performance of Randomized Sample Sort. In fact, the
uniform data distribution used for Figures \ref{fig:det-rand-merge-285}
and \ref{fig:det-rand-merge-tesla} is a \emph{best case} scenario
for Randomized Sample Sort where all bucket sizes are nearly identical.

Figures \ref{fig:det-rand-merge-285}b and \ref{fig:det-rand-merge-tesla}b
show the performance of \textsc{GPU Bucket Sort} for up to $n=256M$
and $n=512M$, respectively. For both architectures, GTX 285 and Tesla
C1060, we observe a very close to linear growth rate in the runtime
of \textsc{GPU Bucket Sort} for the entire range of data sizes. This
is very encouraging for a problem that requires $O(n\log n)$ work.
In comparison with Randomized Sample Sort, the linear curves in Figures
\ref{fig:det-rand-merge-285}b and \ref{fig:det-rand-merge-tesla}b
show that our \textsc{GPU Bucket Sort} method maintains a fixed \emph{sorting
rate} (number of sorted data items per time unit) for the entire range
of data sizes, whereas it is shown in \cite{s7-Sanders} that the
sorting rate for Randomized Sample Sort fluctuates and often starts
to decrease for larger values of $n$.

\section{Conclusions\label{sec:Conclusions}}

In this paper, we presented a \emph{deterministic} sample sort algorithm
for GPUs, called \textsc{GPU Bucket Sort}. Our experimental evaluation
indicates that \textsc{GPU Bucket Sort} is considerably faster than
Thrust Merge \cite{s6-Garland}, the best comparison-based sorting
algorithm for GPUs, and it is exactly as fast as the new Randomized
Sample Sort for GPUs \cite{s7-Sanders} when the input data sets used
are uniformly distributed, which is a \emph{best case} scenario for
Randomized Sample Sort. However, as observed in \cite{s7-Sanders},
the performance of Randomized Sample Sort fluctuates with the input
data distribution whereas \textsc{GPU Bucket Sort} does not show such
fluctuations. In fact, \textsc{GPU Bucket Sort} showed a fixed \emph{sorting
rate} (number of sorted data items per time unit) for the entire range
of data sizes tested (up to $n=512M$ data items). In addition, our
\textsc{GPU Bucket Sort} implementation appears to be more memory
efficient because \textsc{GPU Bucket Sort} is able to sort considerably
larger data sets within the same memory limits of the GPUs.


%

\begin{table*}[tbh]
\centering{}\begin{tabular}{|c|c|c|c|c|}
\hline 
 & Tesla C1060 & GTX 285 (2 GB) & GTX 285 (1 GB) & GTX 260\tabularnewline
\hline
\hline 
Number Of Cores & 240 & 240 & 240 & 216\tabularnewline
\hline 
Core Clock Rate & 602 MHz & 648 MHz & 648 MHz & 576 MHz\tabularnewline
\hline 
Global Memory Size & 4 GB & 2 GB & 1 GB & 896 MB\tabularnewline
\hline 
Memory Clock Rate & 1600 MHz & 2322 MHz & 2484 MHz & 1998 MHz\tabularnewline
\hline 
Memory Bandwidth & 102 GB/sec & 149 GB/sec & 159 GB/sec & 112 GB/sec\tabularnewline
\hline
\end{tabular}\caption{Performance Characteristics For nVIDIA Tesla C1060, GTX 285 with 2
GB memory, GTX 285 with 1 GB memory, and GTX 260. (Source: \cite{tesla,gtx285,gtx260})\label{tab:GPU-Performance-Characteristics}}

\end{table*}

\end{document}